\documentclass{elsart}
\usepackage{natbib}
\usepackage{epsfig}
\begin{document}
\runauthor{Bonitz, Golubnychiy, Filinov, and Lozovik}
\begin{frontmatter}
\title{Single-electron control of Wigner crystallization}
\author[Rostock]{M.~Bonitz,}
\author[Rostock]{V~Golubnychiy,}
\author[Rostock,Moscow]{A.V.~Filinov},
\author[Moscow]{Yu.E.~Lozovik}

\address[Rostock]{Fachbereich Physik, Universit{\"a}t Rostock\\
Universit{\"a}tsplatz 3, D-18051 Rostock, Germany}
\address[Moscow]{Institute of Spectroscopy, 142090 Troitsk, Russia}

%----------------------------------------------------------------------------------------------------
\begin{abstract}
Wigner crystallization in mesoscopic quantum dots containing only
few ($N < 50$) electrons exhibits a number of interesting
peculiarities: (i) there exist two distinct crystal phases, and
(ii) the phase boundary sensitively depends on the precise particle
number. In this paper we demonstrate that this behavior can be
used to control the {\em collective} transport properties by adding or
removing a {\em single electron}.
\end{abstract}
\end{frontmatter}

One of the most exciting properties of mesoscopic systems is the
dependence of their properties on the particle number. Besides the
familiar electron addition spectra in quantum dots, e.g. \cite{ashoori96},
recently, another example has been found,
cf. \cite{bedanov,afilinov-etal.01prl} and references therein: Wigner crystallization
in mesoscopic electron clusters in two-dimensional quantum dots.
Simulations of a small number of electrons confined in a spherically symmetric
harmonic trap and interacting via the Coulomb potential
\cite{afilinov-etal.01prl,afilinov-etal.00pss}
revealed that the
location of the phase boundary of the Wigner
crystal in the density-temperature plane sensitively depends on the precise
particle number. In this paper we discuss principle possibilities of
taking advantage of this behavior for applications.

The particle number sensitivity originates from the configuration 
symmetry of the cluster ground state. Minimization of the total energy 
computed from the hamiltonian 
\begin{eqnarray}
\hat H = -\sum\limits_{i=1}^N \frac{\hbar^2 \nabla_i^2}{2 m^*_i} +
\sum\limits_{i=1}^N \frac{m^*_i \omega_0^2 r_i^2}{2} +
\sum\limits^N_{i<j}\frac{e^2}{\epsilon_b |{\bf r}_{i}-{\bf r}_j|},
\label{Hamil}
\end{eqnarray}
\noindent ($m^{*}$ and $\epsilon_b$ are the effective electron mass and background
dielectric constant, respectively), yields a spherical shell structure details of which 
vary strongly with the particle number, see Fig.~1. ``Magic'' clusters (those with an
integer ratio of particle numbers on the outer and inner shell - e.g. $N=19$) show 
a clear hexagonal crystal structure (Fig.~1, left part). In contrast, non-magic ones 
(e.g. $N=20$)
are dominated by the spherical trap symmetry. This has a strong effect on the stability
of the crystal phases, see Fig.~1, right part.
\begin{figure}[h]
\hspace{-1.3cm}
\centerline{
\psfig{file=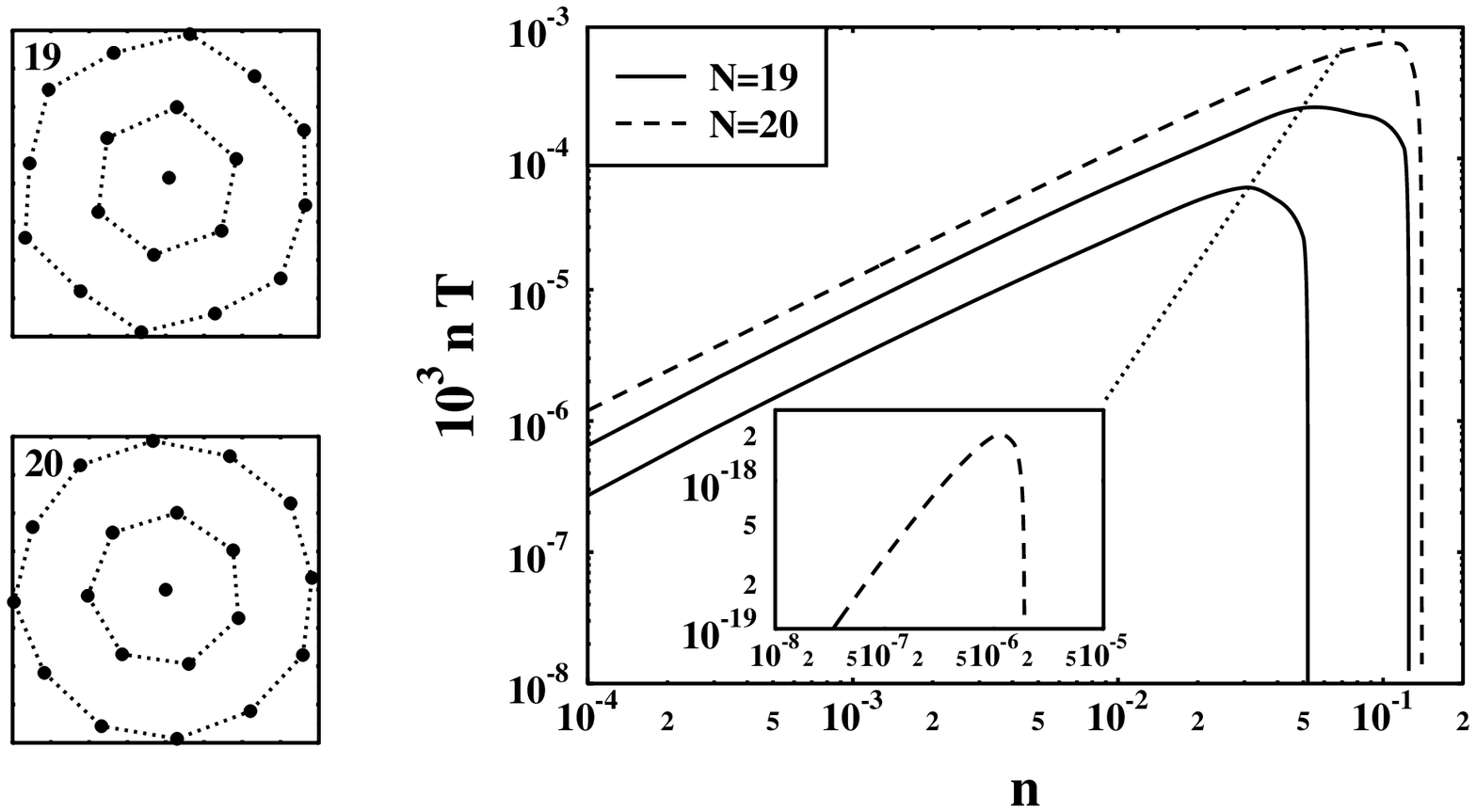,height=14.5cm}}
\vspace{-7.2cm}
\caption[]{\label{phase}
{\em Left two figures}: Ground state configuration of the ``magic'' cluster $N=19$ (top)
and $N=20$ (bottom) in the cluster plane. Each dot is an electron.\\
{\em Right figure}: Phase diagram of the Wigner crystal of 2D clusters with $N=19$ and
$N=20$ electrons. The outer (inner) lines are the radial (angular) melting
phase boundaries. The dotted diagonal line separates the classical (left)
from the quantum (right) crystal. The classical part of the boundaries
is given by lines with constant $\Gamma$, with the values (from top to bottom)
$\Gamma^{RM}_{20}=83, \Gamma^{RM}_{19}=154, \Gamma^{OM}_{19}=330$ and
$\Gamma^{OM}_{20}=3.4 \cdot 10^{11}$. The dimensionless density and
temperature are given by $n=r_s^{-1/2}$ and $T\equiv 1/\Gamma$.
}
\end{figure}

Generally, crystal-like behavior \cite{wc} is found below a critical
temperature (of the order of a few $K$ in semiconductor quantum dots) in
a finite density interval, see Fig.~1. If the density is reduced below a
critical value $n_1(T,N)$, the system undergoes a transition to a state resembling
a classical liquid. Similarly, above a second critical density $n_2(T,N)$
(with $n_2 \ge n_1$) a Fermi liquid-like state (in mesoscopic systems,
sometimes referred to as Wigner molecule) is reached. While this general behavior is
analogous to macroscopic systems, e.g. \cite{tanatar89}, crystallization in
{\em few-electron} systems shows a number of interesting peculiarities \cite{spin}: (i)
 strong $N$ dependence of the phase boundary and (ii), the existence
of a second phase boundary which is embedded into the first one where the crystal
structure is transformed from a completely
ordered (``OO'', fully localized electrons) state into a partially
(``RO'', radially) ordered one where electrons are rigidly confined to crystal shells
which, as a whole, can rotate against each other. The phase boundary is determined
by critical values of the coupling parameters: the low-density boundary by
the classical parameter, $\Gamma\equiv e^2/(\epsilon_b r_0 k_BT)$, and the high-density
limit by $r_s\equiv r_0/a_B$, cf. Fig.~1. Here, $a_B$ is the effective
Bohr radius, $a_B=\hbar^2\epsilon_b/(m e^2)$, and $r_0$ is the mean interparticle distance,
approximately given by the balance of the repulsive Coulomb force and the radial confinement force of
the harmonic potential with strength $\omega_0$: 
$\quad e^2/(\epsilon_b r^2_0)=m\omega_0^2 r_0$.
A particularly strong N dependence is observed for the ``OO'' phase: 
``magic'' clusters ($N=12, 19$ etc.) are found to
be much more stable than clusters having one electron more or less 
\cite{bedanov,afilinov-etal.01prl}.
For example, the orientational melting parameters $\Gamma^{OM}$ and $r_s^{OM}$ of the cluster
with $N=19$ are {\em 9 orders of magnitude} lower than those for $N=20$, see Fig.~1.

This behavior can be exploited for a non-traditional control of crystallization.
In addition to changing temperature or/and density (confinement strength), crystallization
in mesoscopic systems can be achieved by variation of the particle number
{\em without change of $T$ and $n$}. For example, choosing a point in the
temperature-density plane, Fig.~1, which is located between the radial (orientational)
melting curves of $N=19$ and $N=20$ and switching between the two particle
numbers is equivalent to a crossover between crystal-like and liquid-like
(OO and RO crystal) behavior. We will demonstrate this below for the orientational
melting curves in the classical part of the phase diagram, i.e. we fix the
classical coupling parameter $\Gamma$ between
$\Gamma^{OM}_{20}$ and $\Gamma^{OM}_{19}$, (see Fig.~1, right part).
\vspace{6cm}
\begin{figure}[h]
\centerline{
\psfig{file=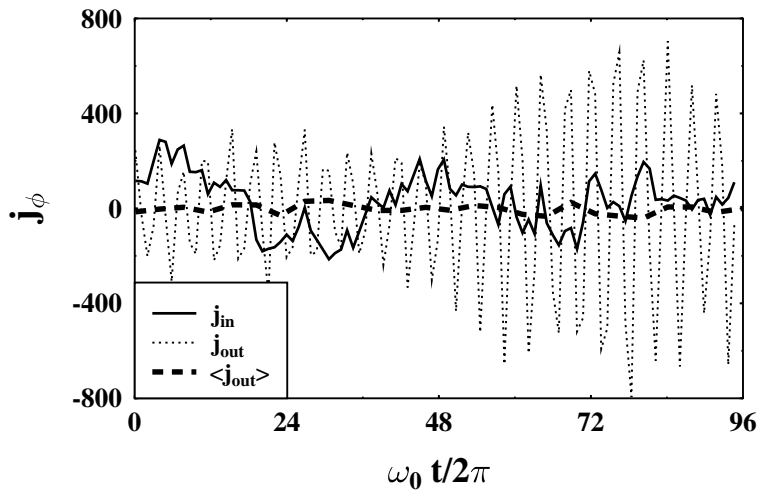,height=8cm}}
\vspace{-7.5cm}
\caption[]{\label{barrier}
Spontaneously excited angular currents of the inner (``in'') and outer (``out'')
shell of an $N=20$ cluster at $\Gamma=500\ll \Gamma_{20}^{OM}$ averaged over $T_0=2\pi/\omega_0$.
The outer shell is pinned. The thick
dashed line is $j_{out}$ averaged over $4 T_0$.
}
\end{figure}
The main difference between the orientationally ordered (with $N=19$) and
disordered ($N=20$) state will be that the latter is able to support inter-shell
rotational excitations. This should give rise to macroscopic currents and magnetic
fields. To estimate this effect, we compute the total
angular current  created by N electrons confined to one of M shells of radius
$R_k$ rotating with angular frequency $\omega_k$ (positive or negative depending
on the direction), $I_{\phi} = e\sum_{k=1}^{M} N_k \, \omega_k R_k$.
The associated magnetic field on the symmetry axis in a distance $z$ above the
cluster plane is directed normal to the plane:
$B_z(z,t)=\frac{\mu \,e}{2}\sum_{k=1}^{M} N_k\,\omega_k(t)
\left[1+\left( z/R_k \right)^2\right]^{-3/2}$
For example, for
$N=20$, we have $N_1=7$, $N_2=12$, $R_2\approx 2 R_1\approx 2\cdot r_0$, so that
$I^{20}_{\phi} \approx e r_0 (7\cdot \omega_1 + 24 \cdot\omega_2)$.
The relation between $\omega_1$ and $\omega_2$ depends on the excitation conditions.
For example, if the total angular
momentum of the excitation is zero, one readily finds $\omega_1/\omega_2 \approx -4N_2/N_1$,
i.e. the inner shell rotates approximately 7 times faster. Alternatively,
rotation of some shells may be inhibited due to defects (pinning). For definiteness,
in the following we consider the cluster $N=20$ with the outer shell pinned.

At a given coupling $\Gamma$ and confinement energy $m\omega_0r_0^2/2$, thermal fluctuations
spontaneously excite rotational and vibrational degrees of freedom both,
in radial and angular direction. The thermal energy
per particle is $k_BT$ and is $\Gamma$ times smaller than the confinement energy. As our simulations
show (see below), approximately half of it is converted into angular kinetic energy (rotations and
oscillations), i.e. $E_{rot} \approx \frac{1}{4}m\omega_0^2R_1^2/\Gamma$. Depending on the
excitation conditions, this gives rise to rotations of the inner
shell electrons with $\omega_{1\,max} \approx \omega_0/\sqrt{2\Gamma}$.
Inserting this result into the expression for $B_z$ and averaging over the fast
vibrations, we obtain
$\displaystyle{
\langle B_{z}(z) \rangle_{max} \approx \frac{\mu \,e^2}{2 \epsilon_b r_0}
\frac{N_1}{\sqrt{m\,r_0\Gamma}}
\left[1+\left( z/r_0 \right)^2\right]^{-3/2}.
}$
Interestingly, the current and the magnetic field increase as the
square root of temperature. In contrast, the cluster of 19 electrons does not support
intershell rotations. Here, the excitation energy is completely converted into vibrations.

We have verified this concept by performing classical molecular dynamics simulations
for $N=19$ and $N=20$ in a wide range of $\Gamma$ values \cite{vova01,anim}.
As expected, the cluster with $N=19$ (pinned to suppress the trivial rotation of the whole
system) supports, at $\Gamma=500>\Gamma^{OM}_{19}$, no intershell
rotations, and only vibrations are excited, even for purely rotational
initial fluctuations \cite{anim}.
In contrast, for $N=20$, strong collective rotational motion of the inner shell electrons
is excited, see Fig.~2. In all cases, we found that, in the spectrum, the vibrational excitations
are well separated from the rotations -- the former occur at significantly higher frequency, 
dominantly at frequency $\omega_0$ (and $\omega_0/4$). After averaging over $4 T_0$, with
$T_0=2\pi/\omega_0$, a ``persistent'' collective current
$\langle j_{in}\rangle$ of the inner shell is observed which exceeds the averaged fluctuating signal
$\langle j_{out}\rangle$ of the pinned outer
shell by at least two orders of magnitude. The direction of the shell rotation and its 
slow time-dependence of $\langle j_{in}\rangle$
are determined by the (random) excitation conditions and is not relevant.
As an independent test
we computed the potential barriers for intershell rotation using Quantum Monte Carlo simulations
\cite{afilinov-etal.01prl}. These simulations effectively average over the random excitations 
and yield, for $N=20$ at $\Gamma=500$, potential barriers which are practically zero. 

For practical applications, one may think of {\em designing} a suitable {\em external}
rotational  excitation
in such a way that it is sufficiently weak so it does not  overcome the barrier of the $N=19$
cluster. If applied to the $N=20$ cluster, the same excitation will easily give rise to
intershell rotations. Such an excitation could be e.g. a constant or pulsed circularly polarized
electric field.
Based on our simulations, we expect that the response of a (pinned) $N=20$ cluster will be a nearly
dissipationless rotation
of the inner shell which stops after removal of one electron \cite{anim}. This allows for a completely
new kind of single-electron controlled devices. Such collective rotations of a group of
electrons can be easily transmitted, e.g. in a system of multiple layers \cite{afilinov-etal.01cpp}
and may find
applications for quantum computing. Naturally, realization of this concept
will require very clean samples, and detection of the weak circular currents is another
important issue to be solved. At the same time, high sensitivity of these few-electron clusters
to very weak rotational excitations could be of interest for
applications by itself.

%\section*{Acknowledgements}
M.B. acknowledges stimulating discussions with participants of NPMS-5, in particular,
J.~Barker on quantum computing and K.~von~Klitzing on experimental verification issues.

This work has been supported by the Deutsche
Forschungsgemeinschaft (grant BO-1366/2) and by a grant for
CPU time at the NIC J\"ulich.

%-----------------------------------------------------------

%\begin{figure}[p]
%\centerline{
%\psfig{file=angbar19.ps,height=9cm,angle=-90}}
%\vspace{-0cm}
%\caption[]{\label{barrier}
%Angular barrier for the cluster of 19 electrons.
%}
%\end{figure}


\begin{thebibliography}{999}
\bibitem{wc} We use the notations ``crystal'', ``melting'' and
``phase transition'' to underline the analogy to the corresponding
phenomena in infinite systems.
\bibitem{spin} In the simulations, the total spin has been fixed. Investigations
on the spin properties of the ground state and of crystallization of mesoscopic clusters are
presently under way.
\end{thebibliography}
\end{document}